\documentclass[10pt]{article} 
\usepackage{ifpdf}
\usepackage{latexsym}
\usepackage{graphics}
\usepackage{epsfig}
\usepackage[dvips]{color}
\usepackage[mathscr]{eucal}
\usepackage{amscd}
\usepackage{amssymb}
\usepackage{amsmath}
\usepackage{amssymb}
\usepackage{amsthm}

\newcommand{\cO}{\mathcal Q}

\newcommand{\lambdai}{\lambda^{(2)}_{(i)}}
\newcommand{\Mone}{M^{(1)}}
\newcommand{\Mtwo}{M^{(2)}}
\newcommand{\Mzero}{M^{\tiny{(0)}}}
\newcommand{\Mthree}{M^{(3)}}
\newcommand{\Alex}{\mathbf I}
\newcommand{\MMI} {\mathbf {M}}
\newcommand{\QQ}{\mathrm Q} 
\newcommand{\e}{\mathbf e}

\newcommand{\emui}{\e^{(k)i}}
\newcommand{\emuj}{\e^{(k)j}}
\newcommand{\Alexflat}{\Alex_0[p,q]}
\newcommand{\volalex}{\mathbf V}
\newcommand{\volflat}{\mathbf {V_0}}
 
\newcommand{\JJ}{\mathbb J}
\newcommand{\bo}{\mathbf r}
\newcommand{\negs}{\!\!\!}

\title{Boundary Term Contribution to the Volume of a  Small Causal Diamond} 
\author{Surbhi Khetrapal${}^a$ and Sumati Surya${}^b$,  \\ ${}^a$ BITS Pilani K. K. Birla Goa Campus
 \\ ${}^b$ Raman Research Institute, Bangalore, India } 

\begin{document}
\maketitle

\begin{abstract} 
  In his calculation of the spacetime volume of a small Alexandrov interval in 4 dimensions Myrheim
  introduced a term which he referred to as a surface integral [1]. The evaluation of this term has
  remained opaque and led subsequent authors to evaluate the volume using other techniques [2]. It
  is the purpose of this work to demystify this integral. We point out that it arises from the
  difference in the flat spacetime volumes of the curved and flat spacetime intervals. An explicit
  evaluation using first order degenerate perturbation theory shows that it adds a dimension
  independent factor to the volume of the flat spacetime interval as the lowest order correction. Our analysis
  admits a simple extension to a more general class of integrals over the same domain.  Using a
  combination of  techniques we also find that the next order correction to the volume
  vanishes.
 \end{abstract}

\section{Introduction} 

Spacetime volume plays a crucial role in the discrete-continuum correspondence of causal set quantum
gravity \cite{blms}. In this approach the fundamental entity underlying the continuum is a locally
finite partially ordered set or causal set.  The order relation of the causal set corresponds to the causal
structure of the spacetime, while  the local finiteness means that underlying every $N$ Planck
volumes of a spacetime region  there are, on average,  $N$ of elements of the causal set.  
Calculations of the volumes of generic but causally well-defined spacetime regions are therefore of
particular interest to causal set theory.

In an insightful CERN preprint on discrete statistical geometry Myrheim \cite{myr} introduced
several key concepts that were subsequently adopted by the causal set approach \cite{blms}. 
An important calculation in this work is of the volume $\volalex$ of a small Alexandrov interval
$\Alex[p,q]$ between two chronologically related events $p$ and $q$. The metric is expanded about
the mid-point $r$ of the flat spacetime geodesic $\gamma_0$ from $p$ to $q$ using Riemann Normal
Coordinates (RNC), with the smallness parameter being the proper time $T$ from $p$ to $q$. In $4$
spacetime dimensions Myrheim obtained an expression for $\volalex$ to order $T^2$ which depends on
the scalar curvature as well as the curvature components along $\gamma_0$.  While the calculation is
straightforward to set up, the evaluation of one of the terms,  referred to as a ``surface
integral''  in \cite{myr} has remained opaque. This led Gibbons and Solodukhin to an evaluation of $\volalex$ which
side steps this integral by instead calculating the universal coefficients that appear in an order
$T^2$ expansion of the volume for the Einstein Static Universe and de Sitter spacetime \cite{gs}.
\begin{figure}[ht] 
\centering \resizebox{2in}{!}{\includegraphics{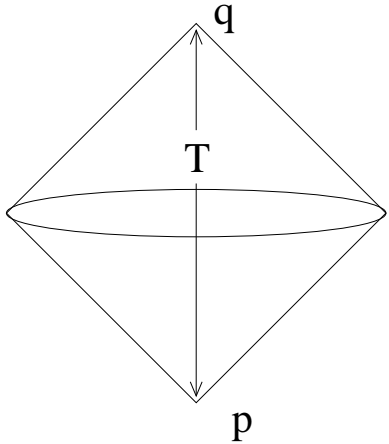}}
\vspace{0.5cm}
\caption{{\small An Alexandrov interval  $\Alex[p,q]$ in flat spacetime. $T$ is the proper time
    between the events $p$ and $q$. }}
\end{figure}

It is the main purpose of this work to decode Myrheim's ``surface'' integral.  Rather than viewing
it as a surface term, we find it more fruitful to think of it as the difference in the {\it flat
  spacetime} volumes of the regions defined by the curved spacetime interval $\Alex[p,q]$ and the
flat spacetime interval $\Alexflat$.  As we will show in Section 2, the boundary of $\Alex[p,q]$ can
be determined to order $T^2$ by merely considering the modification to the light cones in the
tangent spaces $T_pM$ and $ T_qM$ of $p$ and $q$ respectively; the effect of the acceleration of the
null geodesics can be shown to be sub-leading.  The modification in $T_pM$ can be understood from
simple first order degenerate perturbation theory. In the RNC this curvature dependent perturbation
is in fact  restricted to the spatial directions, thus generically transforming the spherical cross section
of the light cone to an ellipsoidal one. Making a coordinate transformation to the principal
directions of this ellipsoid we find an expression for the sum of the modified principal
eigenvalues using first order degenerate perturbation theory. We show that this sum appears
crucially in the Jacobian of transformation relating the flat spacetime integral to Myrheim's term
to lowest order thus completing the first part of our analysis.  In Section 3 we evaluate the
Myrheim term and show that it provides a dimension independent factor to the flat spacetime volume
to this order. This allows us to complete the RNC calculation of the volume $\volalex$ of
$\Alex[p,q]$ up to order $T^2$ for arbitrary $n$.  Our expression agrees with that obtained by
Myrheim \cite{myr} for $n=4$ as well as that obtained by Gibbons and Solodukhin \cite{gs} for
general $n$ using a different technique.  An important by-product of our analysis is that it allows
us to calculate more generic integrals which crop up in causal set theory. We give an example of one
such integral which has recently been used to obtain a new expression for the scalar curvature of a
small causal set \cite{rss} using a curved spacetime generalisation of the work of \cite{meyer}.

What of higher order corrections to the volume? In \cite{lgs} this question was partially addressed
in the discussion on the volumes of large causal diamonds and their asymptotic behaviour.  As
evident from the RNC expansion, higher order corrections to the volume will include higher
derivatives of the curvature. In Section 4 we will examine these terms using the techniques
developed above. While the light cones in $T_pM$ and $T_qM$ can be studied using the next order
perturbation analysis, the acceleration of the null geodesics is non-negligible. Hence it does not
suffice to look at $T_pM$ and $ T_qM$, although it does help determine the types of higher
derivative terms appearing to this order . Using a combination of the analysis of light cones in
$T_pM$, $T_qM$ and the approach of Gibbons and Solodukhin \cite{gs} we evaluate the volume to
$O(T^3)$ in FRW spacetimes. We find that the $O(T^3)$ correction to the volume vanishes altogether.

Because of the appearance of higher derivatives of the curvature  it is tempting to ask if the
volume expansion can have any significance in determining effective actions in quantum gravity.
This idea is not so far-fetched in causal set theory where spacetime volume plays a fundamental
role, making it possible to speculate that higher derivative corrections to the action 
must be determined by these corrections to the volume. This could distinguish the causal set
approach from other approaches to quantum gravity. While
our primary focus in this note is a modest one, namely to decode Myrheim's calculations and to seek
the extension of that analysis, it can be seen as a first step towards this more ambitious goal.
The vanishing of the $O(T^3)$ term then suggests that the corrections to the Einstein-Hilbert action
in the effective continuum action arising from causal set theory will only arise from second
derivatives of the curvature, modulo boundary terms.  However, a systematic approach to prove such a
conjecture is currently beyond our scope.

\section{The Boundary of $\Alex[p,q]$} 

The RNC about a point $\bo \in M$ is defined within a convex normal neighbourhood $\cO$ of $\bo$ in
the spacetime $(M,g)$, i.e., a region $\cO \subset M$ in which the exponential map $\mathrm{exp}: T_p
M \rightarrow \cO $ is a diffeomorphism for any $p \in \cO$. In RNC, the geodesics  emanating from $\bo$ are 
used to coordinatise $\cO$, and the spacetime metric at $\bo$ is taken to be flat, i.e.,
$g_{ab}(\bo)=\eta_{ab}$. This implies that the 
Christoffel connection $\Gamma_{ab}^c(\bo)=0$, so that the metric at any $x \in \cO$ can be expanded as
\begin{equation} g_{ab}(x)=\eta_{ab}(0) - \frac{1}{3} x^c x^d R_{acbd}(0) + O(x^3), \end{equation}
where we have used the RNC identity $\partial_{(b}\Gamma_{ac)}^d(0)=0$. Here, and  in the future we will
often use the short form $0$ to denote the origin   $\bo=(0,0, \ldots 0)$ of the RNC.

The volume of a small Alexandrov interval $\Alex[p,q]$ in $n$ spacetime dimensions
between the points $p=(-T/2,0,0,\ldots,0)$ and $q=(T/2,0,0 ,\ldots 0)$ in RNC is therefore given by
the integral  
\begin{equation}
\label{volintegral} 
 \volalex=\negs \int\limits_{\Alex[p,q]}\negs \sqrt{-g} \,\,d^nx = \int\limits_{\Alex[p,q]}
 \biggl(1-\frac{1}{6}x^cx^dR_{cd}(0) + O(x^3)\biggr) d^nx. 
\end{equation} 
which was first calculated for $n=4$ by Myrheim \cite{myr}.  Importantly, the integration is over a
region $\Alex[p,q]$ which itself is determined by $g_{ab}(x)$ and hence contains corrections to the
flat spacetime interval $\Alexflat$.  Up to $O(x^2)$ this integral can be split into two parts
$\volalex=I_{\MMI} + I_2$,  
where
\begin{equation}  
I_{\MMI}=\negs \int\limits_{\Alex[p,q]} \negs d^nx ,    \qquad I_2=\negs \int\limits_{\Alexflat}
\negs \biggl(  -\frac{1}{6}x^cx^dR_{cd}(0)\biggr) d^nx. 
\end{equation} 
As we will show in the next section, the  second term is straightforward to evaluate in arbitrary
dimensions, much of the simplification arising  from the fact that odd terms do not contribute
to $\Alexflat$ because of its symmetries. 

The integral $I_\MMI$ itself comprises two pieces: the volume $\volflat$ of $\Alexflat$ plus a
contribution $I_{\Delta\Alex[p,q]}$ which Myrheim referred to as a ``boundary'' term. This term was
evaluated for $n=4$ in \cite{myr} without any explanation, although how to evaluate it it has been
far from obvious to subsequent researchers \cite{gs}.  Indeed, it is this precise integral that we
wish to decode in the present work.  We first note that this term is {\it not} a boundary integral
but simply the difference in the {\it flat spacetime} volumes of the interval $\Alex[p,q]$ with
respect to $\eta_{ab}$ and $g_{ab}$ respectively, as realised also in \cite{gs}.

Thus, in order to determine $I_\MMI$, we need to first obtain the boundary of $\Alex[p,q]$ since the
difference in the volumes $I_{\Delta\Alex[p,q]}$ roughly arises from the difference in the range of
integration. In flat spacetime  the boundary  of  $\Alexflat$ is  that of a pair of uniform (light)cones with base radius
$T/2$, one emanating to the past from $q$ and the other to the future from $p$ so that we can
evaluate 
\begin{equation}
\label{flatvol} 
\volflat=\negs \int\limits_{\Alexflat}\negs  d^nx = 2 \int\limits_0^{T/2} \! \!  dt
\int\limits_0^{T/2-t}\negs \!  r^{n-2} dr \int\limits_{S^{n-2}} \! 
d\Omega_{n-2}=\frac{2 A_{n-2}}{n(n-1)}\biggl(\frac{T}{2}\biggr)^n,      
\end{equation} 
where $A_{n-2}$ is the volume of a uniform ${n-2}$ sphere $S^{n-2}$.  

The presence of curvature obviously modifies the boundaries of the two light cones, and it is this
that we now attempt to quantify.  The boundary of $\Alex[p,q]$ is ruled by future-directed and
past-directed null geodesics emanating from $p$ and $q$ respectively.  The tangents to these null
geodesics in turn lie along the future and past light cones in $T_pM$ and $T_qM$ respectively.  The
effect of curvature in general would be to accelerate these null geodesics, so that they no longer
lie along $T_pM$ once they leave $p$, and similarly for $q$. Thus, one would no longer expect as
simple an integration as in Eqn. (\ref{flatvol}).

Nevertheless, the first step to take is to determine the future and past light cones in
$T_{p}M$ and $T_qM$, respectively.  Because of the symmetry of $\Alexflat$ it suffices to restrict
our attention to $T_qM$.  Using the RNC expansion of the metric at $q$
\begin{equation} 
  g_{ab}(q)=\eta_{ab}(0) - \frac{1}{12} T^2R_{0a0b}(0),  \end{equation}   
the tangents $\zeta^a=\frac{dx^a}{ds}=(\zeta^0, \zeta^1, \ldots \zeta^{n-1})$ to the past and future directed null geodesics at $q$ satisfy     
\begin{eqnarray} 
g_{ab}(q)\zeta^a \zeta^b & = & 0   \quad \Rightarrow  \nonumber \\ 
-(\zeta^0)^2+ \sum_{i=1}^{n-1} (\zeta^i)^2 &=& 
\frac{1}{12} T^2R_{0i0j}(0) \zeta^i \zeta^j
\end{eqnarray} 
where we have used the symmetries of the Riemann tensor to simplify
the expression. Thus the  light cone in the tangent space $T_qM$ is given by  the matrix equation   
\begin{equation} 
\label{modnullcone}
{\vec \zeta}^T M {\vec \zeta}=(\zeta^0)^2,
\end{equation}   
where ${\vec \zeta}$ is the spatial part of $\zeta^a$ and 
\begin{equation}
M_{ij}=\delta_{ij} -T^2 \frac{1}{12}R_{0i0j}.  
\end{equation} 
When there is no curvature Eq (\ref{modnullcone}) reduces to  $({\vec \zeta})^2=(\zeta^0)^2$
which is the equation for a uniform  (light)cone, i.e., a cone with $(n-2)$-dimensional spherical cross-sections with 
radii $\zeta^0$. 
  
Since $M$ is a real symmetric matrix, it can be  diagonalised by an orthogonal
matrix $\QQ$, so that inserting $\QQ^T \QQ=1$ in  Eq (\ref{modnullcone})  yields 
\begin{equation} \label{eqellipsoid} 
 {\vec \xi}^T  \Lambda   {\vec \xi} = 
(\zeta^0)^2 \quad  \Rightarrow  \quad 
\sum_i \lambda_i (\xi^i)^2  =  (\zeta^0)^ 2, 
\end{equation} 
where  $ {\vec \xi}= \QQ {\vec \zeta}$ and $\Lambda$ is the diagonalised form of $M$, with
$\Lambda_{ij}=\delta_{ij} \lambda_i$   
This gives an equation for a spatial ellipsoid for  fixed $\zeta^0$ with  principal axes along
the components of $\vec\xi$.  Thus the presence of curvature modifies the regular light cone in
$T_qM$ into one whose spatial sections are ellipsoids with  principal axes given by the $\xi^i$.


A  past-directed null geodesic  from $q$ with tangent vector $\xi^a=(\zeta^0, {\vec \xi})$ satisfying Eq
(\ref{eqellipsoid})  obeys the geodesic equation 
\begin{equation}
\frac{d\xi^a}{ds}|_q=-\Gamma^a_{bc}(q) \xi^b|_q \xi^c|_q =
-\frac{T}{2}\partial_0 \Gamma^a_{bc}(0)  \xi^b|_q \xi^c|_q,      
\end{equation}    
in RNC. In other words, to lowest order, the acceleration of geodesics is zero, as expected.   Expanding along the affine parameter $s$
\begin{equation} 
\label{acceleration} 
\xi^a_s=\xi^a|_q-s\frac{T}{2}  \partial_0
\Gamma^a_{bc}(0) \xi^b|_{q} \xi^c|_{q} + O(s^2), 
\end{equation}  
we see that for small $s$, the second term is sub-leading to order $O(T^2)$ but not to order
$O(T^3)$. Thus, if we find that the first correction to $\xi^a$ is of order $O(T^2)$, then the
acceleration is sub-leading, so that the boundary of $\Alex[p,q]$ is determined by the light cones
in $T_{p,q}M$ alone, to order $T^2$.      

We find the leading order correction to $\xi^a$ by parallel transporting it from $r=(0,0,\ldots, 0)$
to $q$ along a geodesic.  Indeed, as is easily shown, up to order $T^2$ the curve $\gamma=(t, 0,
\ldots,0)$ from $\bo$ to $q$ is a geodesic.  The tangent to $\gamma$ is $U^a=(1,0,\ldots, 0)$, so
that $\partial_0U^b=0$ all along $\gamma$.  For $\gamma$ to be a geodesic therefore
\begin{equation} 
U^a\nabla_a U^b(t)= \partial_0 U^b(t) + \Gamma_{00}^b(t) = t \partial_0\Gamma_{00}^b(0)+O(t^2)
\end{equation}  
must vanish to the required order. 

Using the RNC identity 
\begin{equation} 
\label{rncidentity}
\partial_a \Gamma^d_{bc}(0)  =  -\frac{1}{3}(R_{abc}^{\quad d}(0)+
R_{acb}^{\quad d}(0))
\end{equation} 
and the properties of the curvature tensor, we see that $U^a\nabla_a U^b(t)=O(t^2)$, so that $\gamma$
is a geodesic up to $O(T^2)$.

Now consider the parallel transport of a vector $\omega^a$ from  $\bo$ to $q$ along $\gamma$, $U^b\nabla_b \omega^a=0$ :\begin{equation}
\partial_0 \omega^a + t \partial_0\Gamma_{0b}^a(0) \omega^b=0
\Rightarrow      \partial_0 \omega^a - \frac{t}{3} R_{0b0}^{\quad a}(0)\omega^b=0 
\end{equation} 
where we have used the identity Eq (\ref{rncidentity}).  The symmetries of the Riemann tensor imply
that the time-component of any vector does not change along $\gamma$, i.e., $\partial_0 \xi^0=0$
up to this order.  Since $R_{0j0}^{\quad i}(0)$ is symmetric and
real   (and hence Hermitian) let us use its  eigenfunctions $\{ \emui\}$ 
$R_{0j0}^{\quad i}(0){\emuj}=\rho_{(k)}{\emui}$ to give us a (spatial) orthonormal basis
vectors. Thus the parallel transport equation reduces to  
\begin{eqnarray} 
&& \quad \partial_0 {\emui} - \frac{t}{3} R_{0j0}^{\quad i}(0){\emuj}=0 \nonumber \\ 
&\Rightarrow& \quad
 {\emui(t)}  =  \emui(0)\exp^{\frac{t^2 }{6} \rho_{(k)}}  \nonumber \\ 
&\Rightarrow&  \quad \quad \emui(T/2) \approx  \biggl( \delta^i_j+ \frac{T^2}{24}R_{0j0}^{\quad i}(0) \biggr){\emuj}(0)
\end{eqnarray}     
to this order.  In other words, the spatial components of any vector have an order $O(T^2)$
correction from flat spacetime. Thus, we have shown that the acceleration of the null geodesics at
$q$ is negligible to this order and can therefore be ignored.

We thus find that to $O(T^2)$ the modified past light cone from $q$ is identical to that in $T_qM$
and hence, generically has ellipsoidal instead of spherical sections at constant time. In order to
evaluate the boundary of $\Alex[p,q]$ it therefore suffices to find the relevant properties of this
light cone in $T_qM$, in particular to find the eigenvalues $\lambda_i$ of $M$.

The $\lambda_i$ can be determined to order $T^2$ using first order perturbation theory in $T^2$, or
equivalently, second order perturbation in $T$ where the first order perturbation is zero. We
prefer to use the latter terminology since it will make it easier to adapt to the $O(T^3)$
perturbations.   
Rewriting 
\begin{equation}   
M=\Mzero +T^2 \Mtwo 
\end{equation} 
where $\Mzero=I$ and $\Mtwo_{ij}=- \frac{1}{12}R_{0i0j}$, we note that the zeroth order
eigenfunctions are simply the unit vectors $\psi^{(0)}_{(i)} $ in the principal directions, with
eigenvalues $\lambda^{(0)}_{(i)}=1$. The first correction $\lambda^{(2)}_{(i)}$ to the
$\lambda^{(0)}_{(i)}$ are then the eigenvalues of the operator ${\psi^{(0)}_{(i)}}^T \Mtwo
\psi^{(0)}_{(j)}$, which in this case is simply $\Mtwo$ itself.

It is in fact not necessary for us to solve the exact eigenvalue problem in order to obtain an
expression for the volume up to this order. As will become apparent shortly, all that is required is
to obtain the sum $\sum_i {\lambda^{(2)}_{(i)}}$. The characteristic equation $||\Mtwo-\lambda I ||
=0 $ 
gives rise to  an $(n-1)$th order polynomial
\begin{equation} 
 \lambda^{n-1} - \lambda^{n-2} \sum \Mtwo_{ii} + O( \lambda^{n-3})= 0.  
\end{equation} 
Rewriting this equation in terms of its roots $\lambdai$ 
\begin{equation} \prod\limits_i^{n-1}\biggl(\lambda- \lambdai\biggr)= 0, \end{equation} 
we see that  the coefficient  of the $(n-2)$th order term is 
\begin{equation} 
\label{sumev} 
-\sum_i \lambdai = -\sum_i \Mtwo_{ii}=\sum_i \frac{1}{12}R_{0i0i}=\frac{1}{12}R_{00} . 
\end{equation} 

\begin{figure}[ht] 
\centering \resizebox{2in}{!}{\includegraphics{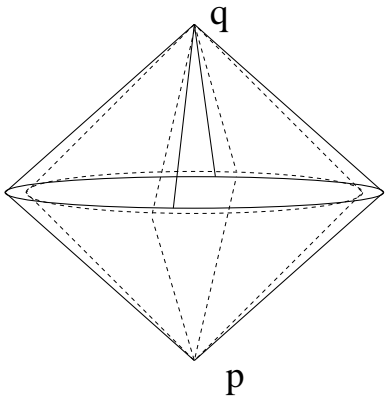}}
\vspace{0.5cm}
\caption{{\small The effect of curvature on the boundary of $\Alex[p,q]$ up to order $T^2$. The
    dotted lines indicate the symmetric  boundary with respect to flat spacetime and the bold lines
    the modified boundary in the presence of curvature. In this figure the two principal directions
    are taken to be non-degenerate.}} \label{curvedalex.fig}
\end{figure}

While we will not require for our main purpose to solve for $\lambdai$, it is nevertheless
instructive to see the effect of curvature on the light cones in a simple example.  For $n=3$
\begin{equation} 
\lambda^{(2)}_{\pm}  =\frac{1}{2}\biggl(\Mtwo_{11}+\Mtwo_{22} \pm \sqrt{ (\Mtwo_{11}-\Mtwo_{22})^2+ 4
  (\Mtwo_{12})^2}\biggr) . 
\end{equation} 
Thus, the light cone in $T_qM$ is transformed from a symmetric cone to one with elliptical sections
at constant $\zeta^0$ with the lengths of the semi-major and semi-minor axes being
${\zeta^0}/{\sqrt{\lambda_{-}}}$ and ${\zeta^0}/{\sqrt{\lambda_{+}}}$, respectively.  The principal
directions are degenerate, i.e., $\lambda^{(2)}_{+}=\lambda^{(2)}_{-} $ iff $\Mtwo_{11}=\Mtwo_{22}$
and $\Mtwo_{12}=0$, i.e., the curvature is isotropic.  In this case $\lambda_{\pm}=1-\frac{1}{24}
T^2R_{00}$, which means that the light cone, though still symmetric, is narrower than in flat
spacetime if $R_{00}>0$ and wider otherwise.  We can expect a straightforward generalisation of this
behaviour in higher dimensions -- the light cone could be wider in some spatial directions and
narrower in others depending on the curvature.  These various possibilities are already evident in
$n=3$. For example, if $\sum_i\Mtwo_{ii} = -\frac{1}{12}R_{00}$ is equal to $ \Delta =
((\Mtwo_{11}-\Mtwo_{22})^2+ 4 (\Mtwo_{12})^2)^{1/2} $ (i.e., $M_{12}=\sqrt{M_{11}M_{22}}$ ),
$\lambda_+<1$ for $R_{00}>0$ and $\lambda_+>1$ for $R_{00}<0$ while $\lambda_-=1$ in both cases. If
$R_{00}>0$ and $\Delta>\frac{1}{12}R_{00}$, then $\lambda_+>1$ and $\lambda_-<1$, etc.  Figure
\ref{curvedalex.fig} shows one of these possibilities.

\section{The Volume Calculation} 

We are now in a position to calculate the volume of $\Alex[p,q]$ in arbitrary dimensions. But first
let us compare this approach with the one in \cite{gs}. The starting point in \cite{gs} was to
assume a form for the volume  
\begin{equation}
\volalex=\volflat(1+ \alpha(n)R(0)T^2 + \beta(n) R_{00}(0) T^2 + O(T^3)),  
\end{equation}  
based on Myrheim's calculation, thus bypassing the explicit evaluation of Eq
(\ref{volintegral}). The universal constant $\alpha(n)$ was evaluated by calculating the volume of a
causal diamond explicitly for the Einstein static universe for which $R_{00}$ is identically zero,
and expanding in powers of $T$. Using this, the universal constant $\beta(n)$ was calculated from
the volume of a causal diamond de Sitter spacetime ($R,R_{00} \neq 0$), again by expanding in powers
of $T$.  Thus, the authors of \cite{gs} obtained a general formula for the volume of a small causal
diamond in $n$ spacetime dimensions to order $T^2$, and this will serve as a check for our
calculations. Moreover, this technique of \cite{gs} will prove to be useful in our exploration of
the next higher order corrections to the volume in Section 4.

As shown in the previous section, the contribution from $I_\MMI$ to the volume to $O(T^2)$
comes from a region bounded by light-cones whose tangent vectors at $q$ satisfy Eq
(\ref{eqellipsoid}). The boundary of the backward light-cone from $q$ is therefore given by
\begin{equation}
\biggl(\frac{T}{2} -t\biggr)^2= \sum_i \lambda_i y_i^2, 
\end{equation}  
where we have chosen the spatial coordinates $y_i$ along the principal axes of the spatial
ellipsoids. We may therefore construct the nested integral   
\begin{equation}
\label{IMMI} 
I_{\MMI}= 2 \int\limits_{0}^{T/2} dz_0 \int\limits_{-w_0}^{w_0} dy_1
\int\limits_{-w_1}^{w_1}dy_2 \ldots \int\limits_{-w_{(n-2)}}^{{w_{(n-2)}}} dy_{n-1}
\end{equation}   
where $z_0=T/2-t$, $ {\sqrt{\lambda_{k+1}}}w_k=\sqrt{z_0^2-\sum_{i=1}^{k}\lambda_{(i)} y_i^2}$.  The
coordinate transformation $z_i=\sqrt{\lambda_i} y_i$ simplifies the integral to that of a
regular(uniform) cone along with a factor corresponding to the Jacobian of transformation
\begin{equation} 
\label{jacobian} 
\JJ= \frac{1}{\sqrt{\lambda_1\lambda_2 \ldots \lambda_{(n-1)}}} = 1 -
\frac{T^2}{2}\sum\limits_{i=0}^{n-1} \lambdai +O(T^3), 
\end{equation} 
since $\lambda_i=1+T^2\lambdai+O(T^3)$. Thus, 
\begin{equation} 
I_{\MMI}=\volflat\biggl(1-\frac{T^2}{2}\sum_{i=1}^{n-1} \lambdai\biggr)
=\volflat\biggl(1+\frac{T^2}{24}R_{00}(0)\biggr)  
\end{equation} 
where we have used Eq (\ref{sumev}). Thus, our main result is that  in any  dimension the boundary
term of Myrheim  \cite{myr}  takes the form 
\begin{equation} 
I_{\Delta\Alex[p,q]}=\volflat\, \biggl( \frac{T^2}{24}\, R_{00}(0)\biggr). 
\end{equation}

We complete this section by including the computation of the $I_2$ integral in arbitrary
dimensions. Let us rewrite 
\begin{equation}
 I_2=\int\limits_{\Alexflat} \biggl( -\frac{1}{6}x^cx^dR_{cd}(0)\biggr) d^nx=-\frac{1}{6}(J_1+J_2+J_3)
\end{equation} 
where 
\begin{equation} 
J_1 = \negs \! \! \int\limits_{\Alexflat}\! \! \! d^nx\,\,  t^2\,R_{00}(0), \, \,  J_2  =   2 \sum\limits_{i=1}^{n-1}\int\limits_
{\Alexflat} \! \! \! d^nx \, \, t  \, x^i \,  R_{0i}(0),  \,\,
J_3 =  \negs \sum\limits_{i,j=1}^{n-1}\int\limits_ {\Alexflat}\! \! \!  d^nx \, \, x^ix^j\,R_{ij}(0).
\end{equation} 
Evaluating 
\begin{eqnarray} 
J_1 & = & 2 R_{00}(0) \int\limits_{0}^{\frac{T}{2}} dt \, \, t^2 \int\limits_{0}^{\frac{T}{2}-t} dr
\, \, r^{n-2} \int\limits_{S^{n-2}}
d\Omega_{n-2} \nonumber \\ 
& = & 
\frac{2 A_{n-2}}{n-1} \, \,R_{00}(0) \int\limits_{0}^{\frac{T}{2}} dt\, \, t^2 \biggl(\frac{T}{2} -t\biggr)^{n-1}  \nonumber \\ 
& = &  \frac{4 A_{n-2}}{n(n-1)(n+1)(n+2)} \,\,  R_{00}(0)   \biggl(\frac{T}{2}\biggr)^{n+2}.  
\end{eqnarray} 
To evaluate $J_2,J_3$ it is convenient to transform from the Cartesian coordinates
$x^i$ to spherical polar coordinates $x^i=r f^i(\Omega)$, where
\[  
f^i(\theta_1, \theta_2 \ldots \theta_{n-2}) =
\left\{
              \begin{array}{ll}
                   \prod\limits_{k=1}^{n-i-1}\sin\theta_k \cos\theta_{n-i}  & (i>1)\\ \\ 
                   \prod\limits_{k=1}^{n-2}\sin\theta_k & (i=1)
              \end{array}
\right.
\]
Since the  angular contribution to $J_2$ is odd, it must  therefore vanish.  Similarly, the
only contribution to $J_3$ comes from the even terms. The contribution from each $x^i$ being the
same,  by spatial spherical symmetry of $\Alexflat$ , we may  simply
evaluate the integral for $x^{n-1}=r\cos \theta_1$ 
\begin{eqnarray}    
J_3& = & \sum_{i=1}^{n-1}\int\limits_ {\Alexflat} \! \! d^nx \, \, (x^i)^2R_{ii}(0) \nonumber \\
&=&2 \biggl(\sum_{i=1}^{n-1}R_{ii}(0)\biggr)  \int\limits_{0}^{\frac{T}{2}} dt \int\limits_0^{\frac{T}{2}-t} dr \,\, r^n\int\limits_{S^{n-2}} d\Omega_{n-2}
\cos^2 \theta_1  
\end{eqnarray} 
where 
\begin{eqnarray} 
\int\limits_{S^{n-2}} d\Omega_{n-2} \cos^2 \theta_1 & =&  \int d\theta_1 \ldots d \theta_{n-2} \biggl(\prod_{i=1}^{n-3}
\sin^{(n-i-2)}\theta_i\biggr) \cos^2 \theta_1 \nonumber \\ 
&=&  \int\limits_{0}^\pi d\theta_1 \sin^{n-3}  \theta_1 \cos^2 \theta_1  \int\limits_0^\pi d \theta_{2} \sin^{n-4}
\theta_2  \ldots \int\limits_0^{2\pi} d\theta_{n-2} \nonumber \\
&=& \frac{A_{n-2}}{n-1},  
\end{eqnarray} 
so that  
\begin{equation} 
J_3 = \frac {2 A_{n-2}} {(n-1)(n+1)(n+2)} \biggl(\sum_{i=1}^{n-1} R_{ii} \biggr) \biggl(\frac{T}{2}\biggr)^{n+2} 
\end{equation} 
Thus 
\begin{eqnarray} 
I_2 &= & -\frac{1}{3}\frac{A_{n-2}}{(n-1)(n+1)(n+2)} \biggl(\frac{T}{2}\biggr)^{n+2} \biggl(\biggl(\frac{2}{n}+1\biggr)R_{00}(0)+
R(0)\biggr) \nonumber \\ 
& = &  -\frac{n \volflat}{24(n+1)(n+2)} T^2 \biggl(\biggl(\frac{2}{n}+1\biggr)R_{00}(0)+ R(0)\biggr)
\end{eqnarray} 
Adding this to $I_{MMI}$ we get 
\begin{equation} 
\label{volgenn} 
\volalex=\volflat\biggl( 1 + \frac{n}{24(n+1)} T^2 R_{00}(0)-\frac{n}{24(n+1)(n+2)}T^2 R(0)\biggr) 
\end{equation}  
which matches Myrheim's expression for $n=4$ \cite{myr} as well as the expression in \cite{gs}
for arbitrary $n$.

As discussed in the introduction, an important reason to be able to calculate the integral $I_\MMI$
explicitly is so that we can extend the analysis to other integrals that make their appearance in
causal set theory.  In \cite{rss}, for example, the following integral must be evaluated for
integer $m\geq 0$:  
\begin{equation}
K=\int\limits_{\Alex[p,q]} d^ny \,\,\tau^m,  
\end{equation} 
where $\tau$ is the proper time from the event that one is integrating over and the future most
point $q$ of $\Alex[p,q]$. Since the boundary of the light cones are given by Eq
(\ref{eqellipsoid}), we see that $\tau^2=(T/2-t)^2-\sum_i\lambda_i y_i^2$. As in the evaluation of
$I_\MMI$ the limits of the integrals are again given as in Eq (\ref{IMMI}). Again, performing a
change of coordinates $z_i=\sqrt{\lambda_i} y_i$ simplifies the integral to
\begin{equation} 
K=\frac{1}{\sqrt{\prod_i\lambda_i}}\int\limits_{\Alexflat} d^nz \,\, \tau^m= \biggl(1+
  \frac{T^2}{24} R_{00}(0)\biggr) \int\limits_{\Alexflat} d^nz \,\, \tau^m. 
\end{equation} 
The universality of this term proves to be  crucial in determining a recurrence formula for 
the average numbers of ``$k$-chains''  in  a causal set $C$ which is approximated by  $\Alex[p,q]$. This
in turn allows us to find an expression for the scalar curvature as well as the dimension in purely
order theoretic terms \cite{rss}.


\section{On the Order $T^3$  Corrections to the Volume}

In this section we show that the next higher order correction $O(T^3)$ to the volume is in fact
zero. 

To order $T^3$, the volume is given by 
\begin{equation} 
\volalex \! =\negs \int\limits_{\Alex[p,q]} \! \! \! \sqrt{-g} \,\,d^nx = \negs \int\limits_{\Alex[p,q]}
 \! \! \! \biggl(1-\frac{1}{6}x^cx^dR_{cd}(0) -\frac{1}{12} x^ex^cx^d\partial_eR_{cd}(0)+O(x^4)\biggr) d^nx. 
\end{equation} 
where we have used the $O(x^3)$ expansion in the RNC 
\begin{equation}
g_{ab}(x)=\eta_{ab}(0) - \frac{1}{3} x^c x^d R_{acbd}(0) -\frac{1}{6}  x^ex^cx^d\partial_eR_{acbd}(0) + O(x^4)
\end{equation} 
which arises from the identity $\partial_{(a}\partial_b \Gamma^e_{cd)}(0) = 0 $ or
equivalently
\begin{equation} 
\label{identitytwo} 
\partial_d \partial_{(e} \Gamma^a_{b)c}(0)  = - \frac{1}{6} \biggl( 2 \partial_{(e} {R^a}_{b)cd}(0)
+ \partial_e {R^a}_{cbd}(0) + \partial_d {R^a}_{ecb}(0)\biggr) .
\end{equation}

Again, we may split up the above integral into a piece that comes entirely from the flat spacetime
interval $\Alexflat$ and a ``remainder'' coming from the difference in the two integrals
\begin{eqnarray} 
\volalex &= &  I_1 + I_2+I_3 \nonumber \\ 
I_1& = & \int\limits_{\Alex[p,q]} d^nx  \nonumber \\ 
I_2&=& -\int\limits_{\Alex[p,q]} \frac{1}{6}x^cx^dR_{cd}(0)
d^nx,     \nonumber \\ 
I_3 &=& \int\limits_{\Alexflat}
 \biggl(-\frac{1}{12} x^ex^cx^d\partial_eR_{cd} +O(x^4) \biggr)
 d^nx. 
 \end{eqnarray} 
 The symmetry of $\Alexflat$ means that the leading order contribution to $I_3$, being odd,
 vanishes.  In addition, since the leading order correction to $\Alex[p,q]$ is of order $T^2$ and
 the integrand of $I_2$ is $O(x^2)$, this term must be $O(T^4)$. Hence the only $O(x^3)$
 contribution can come from $I_1$ which is analogous to the Myrheim integral.

The $O(T^3)$ corrections to $\volalex$ should be of the form $V^aW^bZ^c \partial_a R_{bc} (0)$ where
the $V^a,W^a, Z^c$ can be time-like or space-like vectors. Since to $O(T^2)$ one only has terms of
$R$ and $R_{00}$, a first guess is to include terms like $\partial_0 R$ and  $\partial_0 R_{00}$.
We show that these are precisely the sorts of terms that appear when examining the light-cones in
$T_pM$ and  $T_qM$ to this order.  

To order $T^3$, the metric at $q$ is 
\begin{equation}
g_{ab}(q)= \eta_{ab}(0) - \frac{1}{12} T^2R_{0a0b}(0) -\frac{1}{48} T^3\partial_0 R_{0a0b}(0) ,
\end{equation} 
so that for $\zeta^a \in T_qM$  satisfying the null condition  $g_{ab}(q)\zeta^a \zeta^b  =  0$   
\begin{equation}  
 -(\zeta^0)^2+ \sum_{i=1}^{n-1} (\zeta^i)^2 =  
\biggl( \frac{1}{12} T^2R_{0i0j}(0) 
+ \frac{1}{48} T^3\partial_0 R_{0i0j}(0) \biggr) \zeta^i \zeta^j,  
\end{equation} 
or ${\vec \zeta}^T M {\vec \zeta}=(\zeta^0)^2$ which is the next order modification of Eq
(\ref{modnullcone}), where now
\begin{equation}
M_{ij}=\delta_{ij} - \frac{1}{12}T^2 R_{0i0j}-\frac{1}{48}T^3\partial_0 R_{0i0j} = \Mzero_{ij} +
T^2\Mtwo_{ij} + T^3\Mthree_{ij}.
\end{equation} 
This suggests that only the time-derivatives of the Riemann tensor is relevant to the volume
calculation. Moreover, 
using Eqn. (\ref{identitytwo}) we see that the curve $\gamma$ from $r $ to $q$ along the $t$-axis is
also  a  geodesic to $O(T^3)$. This follows from the fact that
\begin{equation} 
U^a\nabla_a U^b= \frac{t^2}{2} \partial_0^2  \Gamma_{00}^b(0)
+ O(t^3)=O(t^3) , 
\end{equation} 
since $\partial_0^2\Gamma_{0a}^b(0)=0$ from (\ref{identitytwo}). Moreover, again to this order the
time-component of any vector remains unchanged under parallel transport  while the spatial components
satisfy
\begin{equation}
\partial_0 {\emui} - \frac{t}{3} R_{0j0}^{\quad i}(0){\emuj}-\frac{t^2}{6} \partial_0 R_{0j0}^{\quad
  i}(0){\emuj}=0. 
\end{equation}     
Thus, at least in $T_qM$, even to order $T^3$, the only derivative of the curvature that appears is
the time derivative.  If we were to ignore the acceleration of the null geodesics, and use  third order perturbation
theory to find  the eigenvalues of $M$, then again, $I_1$ differs from  the flat space-integral
purely by the Jacobian $\JJ$ Eq (\ref{jacobian}), which to this order is
\begin{equation} 
\JJ= 1- \frac{T^2}{2} \sum\limits_{i=1}^{n-1} \lambdai - T^3\sum\limits_{i=1}^{n-1} \lambda_i^{(3)} . 
\end{equation}  
Given the form of $\Mthree$, this suggests that only the trace $\sum\limits_{i=1}^{n-1}\partial_0
R_{0i0j} = \partial_0 R_{00}$ will contribute to $\JJ$.  We show that this is indeed the case, using
a simple extension of standard third order perturbation theory.  Expanding $M$, its eigenvalues and
its eigenfunctions in powers of $T$
\begin{eqnarray} 
&& \biggl( \Mzero + T\Mone + T^2 \Mtwo + T^3 \Mthree + \ldots \biggr) \biggl( \psi_0+ T\psi_1 + T^2
\psi_2 + T^3 \psi_3 \biggr)  \nonumber \\ 
&= &   \biggl( \lambda_0 + T\lambda_1 + T^2 \lambda_2 + T^3 \lambda_3 + O(T^4) \biggr)
\biggl( \psi_0+ T\psi_1 + T^2 \psi_2 + T^3 \psi_3 \biggr)
\end{eqnarray} 
gives the following set of equations 
\begin{eqnarray} 
(\Mzero-\lambda_0)\psi_0 &=&0 \nonumber \\ 
(\Mzero-\lambda_0)\psi_1 &=&(\lambda_1-\Mone)\psi_0  \nonumber \\ 
(\Mzero-\lambda_0)\psi_2 &=&(\lambda_1-\Mone)\psi_1+(\lambda_2-\Mtwo)\psi_0   \nonumber \\ 
(\Mzero-\lambda_0)\psi_3 &=&(\lambda_1-\Mone)\psi_2 + (\lambda_2-\Mtwo)\psi_1 +
(\lambda_3-\Mthree)\psi_0 .  
\end{eqnarray} 
Because $\Mone=0$, the second equation tells us that $\lambda_1=0$. This in turn means that $\psi_1$
is an eigenfunction of $\Mzero$.  Using the freedom to add any multiple of $\psi_0$ to $\psi_s$,
$s>0$, we can arrange $(\psi_0,\psi_s)=0$ for all $s>0$ \cite{schiff}. In particular, we may choose
$\psi_1=0$.  Hence, contracting the last equation with $\psi_0$ gives us
\begin{equation}
\lambda_3 \langle\psi_0 |\psi_0\rangle =\langle\psi_0 | \Mthree| \psi_0\rangle. 
\end{equation}  
Going back to our notation 
\begin{equation}
 \sum\limits_{i=1}^{n-1} \lambda_i^{(3)} = \sum\limits_{i=0}^{n-1} \Mthree_{ii}  = -
 \frac{1}{48} \partial_0 R_{00}. 
\end{equation}

However, as discussed earlier, the acceleration of a null-geodesics at $q$, though sub-leading in
$O(T^2)$ is not sub-leading to $O(T^3)$ and hence cannot be ignored.  It is at present not clear to
us how to evaluate this contribution to the volume due to the acceleration.  Instead, we resort to
the Gibbons-Solodukhin approach by calculating the volumes of small causal diamonds in FRW
spacetimes for which $\partial_0 R$ and $\partial_0 R_{00} $ do not vanish.

Based on the considerations above we take the $O(T^3)$ correction to the volume to be of the form
\begin{equation}    
\volflat\biggl( \chi(n)\partial_0
R(0) + \kappa(n) \partial_0 R_{00}(0)\biggr),   
\end{equation} 
where again $\chi(n)$ and $\kappa(n)$ are universal constants to be evaluated.  A more covariant
version of the order $T^3$ terms is $T^a T^2 \partial_a R(0)$ and $ T^a T^b T^c \partial_a R_{bc}(0)$
which reduces to the above  when the proper time is aligned along the time axis.

We now calculate $\chi(n)$ and $\kappa(n)$ by finding the volume of a small diamond in
$n$-dimensional  FRW
spacetimes.   
It suffices to consider a spatially flat class of FRW spacetimes
\begin{equation}
ds^2 = - dt^2 + t^{2\sigma}  \sum\limits_{i=1}^{n-1} (dx^i)^2, 
\end{equation} 
where we have taken the scale factor to be of the form $a(t) = t^\sigma $.  The Alexandrov interval
we will evaluate is centered at some $t=t_0$, with $p$ and $q$ chosen appropriately.  We will
examine two special cases $\sigma=1$ and
$\sigma=1/2$. \\

\noindent {\bf Case I: $\sigma=1$}. \\ 
  
We will work in  conformal time,  $\eta =   \ln(\frac{t}{t_0})$ or $t=t_0 \exp^{\eta}$ , where we
have chosen the integration constant so that $\eta = 0$ at $t=t_0$, and $t_0 >  0$. 
\begin{equation}
ds^2 = \tilde{a}(\eta)^2 \biggl(- d\eta^2 + \sum\limits_{i=1}^{n-1} (dx^i)^2\biggr) 
\end{equation}
where  $\tilde{a}(\eta) =  t_0 e^{\eta}$. For this spacetime 
\begin{eqnarray} 
R(t)=\frac{(n-1)(n-2)}{t^2}, & \quad &R_{00}=0 \nonumber \\ 
\Rightarrow \partial_0 R(t)=-\frac{2(n-1)(n-2)}{t^3}, &\quad &\partial_0 R_{00}=0,
\end{eqnarray} 
which allows us to calculate $\chi(n)$.     

We now calculate the volume of an interval  $\Alex[p,q]$ where in conformal coordinates  $p =
(-\frac{N}{2}, 0, \ldots, 0)$ and  $q = (\frac{N}{2}, 0, \ldots, 0)$, so that the proper time $T$ between $p$ and $q$ is:
\begin{equation} 
T  = t_q - t_p = t_0 (e^{\frac{N}{2}} - e^{-\frac{N}{2}})  = 2 t_0 \sinh(\frac{N}{2}), 
\end{equation} 
or 
\begin{equation}
N  = 2 \sinh^{-1}(\frac{T}{2t_0})  = \frac{T}{t_0} (1 - \frac{1}{24} {\frac{T}{t_0}}^2 + O(T^4)). 
\end{equation}
Since $g_{ab}$ is conformally flat the boundary of $\Alex[p,q]$ is determined by  the equation $r =
\frac{N}{2} - |\eta|$. 
\begin{eqnarray} 
V(\tau) \negs & = & \negs  V_++ V_- \nonumber \\ 
&=& \negs \int \limits_{0}^{\frac{N}{2}} \! \!  d\eta \, \, \tilde{a}(\eta)^{n} \! \! \int \limits_0^{\frac{N}{2}-\eta}
\! \!  dr \, \,  r^{n-2} \! \! \int\limits_{S^{n-2}} \! \!  \! d \Omega_{n-2} +\int \limits_{-\frac{N}{2}}^0
\! \! d\eta \, \, \tilde{a}(\eta)^{n}
\! \! \int \limits_0^{\frac{N}{2}+\eta} \! \!  dr \, \,  r^{n-2} \! \! \! \int\limits_{S^{n-2}} \! \! d \Omega_{n-2} \nonumber \\
& = & \negs \frac{2 A_{n-2}}{n-1}t_0^n \int\limits_0^{\frac{N}{2}}
\cosh(n\eta)\biggl(\frac{N}{2}-\eta\biggr)^{n-1} 
\end{eqnarray} 

The integral 
\begin{eqnarray} 
I &=& \int\limits_0^{\frac{N}{2}} \cosh(n\eta)\biggl(\frac{N}{2}-\eta\biggr)^{n-1}  \nonumber  \\ 
&=&  \sum\limits_{k=0}^\infty \sum\limits_{l=0}^{n-1} \frac{n^{2k}}{(2k)!}
\biggl(\frac{N}{2}\biggr)^{n-1-l} (-1)^l \binom{n-1}{l} \int\limits_0^{\frac{N}{2}}d\eta \,
\eta^{2k+l} \nonumber \\ 
&=& \biggl( \frac{N}{2} \biggr)^n \frac{1}{n} \biggl( 1+ \frac{n^2N^2}{4(n+1)(n+2)}+O(N^4)\biggr),    
\end{eqnarray} 
where we have used the relation 
\begin{equation}
\label{binomreln} 
\sum_{l=0}^n \frac{(-1)^l}{(l+a)} \binom{n-1}{l} = \frac{1}{a}\binom{n-1+a}{n-1}^{-1}. 
\end{equation}
Using the expansion 
\begin{equation} 
\biggl(\frac{t_0 N}{2}\biggr)^{n} = \biggl(\frac{T}{2}\biggr)^{n}\biggl(1 - \frac{n}{24} \biggl(\frac{T}{t_0}\biggr)^2
+O(T^4) \biggr), 
\end{equation}
we find the expression for the volume 
\begin{eqnarray} 
\volalex & = & \volflat  \biggl( 1 -  T^2\frac{n(n-1)(n-2)}{24(n+1)(n+2)} \frac{1}{t_0^2} + O(T^4)
  \biggr)  \nonumber \\ 
& = & \volflat \biggl( 1 - T^2 \frac{n}{24(n+1)(n+2)} R(t_0)   + O(T^4) \biggr) 
\end{eqnarray} 
Clearly, even though $\partial_0 R(t_0) = -2 (n-1)(n-2) {t_0}^{-3} \neq 0$, the $O(T^3)$ terms vanish,
hence implying that $\chi(n)=0$.  \\ 
   
\noindent {\bf Case II: $\sigma=1/2$}. \\ 

Conformal time in this case is given by $\eta = 2(\sqrt{t} -\sqrt{t_0})$, where again  $\eta = 0$ at
$t=t_0$, $t_0 > 0$, so that  $t = t_0 \biggl(1+\frac{\eta}{2\sqrt{t_0}}\biggr)^2$.  The conformal
time scale factor  is therefore $\tilde{a}(\eta) =  \sqrt{t_0} + \frac{\eta}{2}$.  For this
spacetime 
\begin{eqnarray} 
R(t)  =  \frac{(n-1)(n-4)}{4t^2},  &\quad&   R_{00}(t)=\frac{(n-1)}{4t^2}  \nonumber \\ 
\Rightarrow \partial_0 R(t)=  -\frac{(n-1)(n-4)}{2t^3}, &\quad & \partial_0R_{00}(t) =-
\frac{(n-1)}{2t^3}. 
\end{eqnarray} 
Since we have already shown that $\chi(n)=0$, we can use this to calculate $\kappa(n)$.   
For $p,q$ given as before,  the proper time  $T$ from $p$ to $q$ is:
\begin{equation}
T = t_q - t_p  =  \sqrt{t_0} N,  \Rightarrow N = \sqrt{t_0}^{-1} T 
\end{equation}
Thus, 
\begin{eqnarray} 
\volalex \negs & = & \negs  V_++V_- \nonumber \\ 
&=& \negs \int \limits_{0}^{\frac{N}{2}} d\eta \, \, \biggl( \sqrt{t_0}+ \frac{\eta}{2} \biggr)^{n} \int \limits_0^{\frac{N}{2}-\eta}
dr \, \,  r^{n-2} \int\limits_{S^{n-2}} d \Omega_{n-2}  \nonumber \\ 
&& +\int \limits_{-\frac{N}{2}}^0 d\eta \, \,  \biggl( \sqrt{t_0}+ \frac{\eta}{2} \biggr)^n
\int \limits_0^{\frac{N}{2}+\eta} dr \, \,  r^{n-2} \int\limits_{S^{n-2}} d \Omega_{n-2} \nonumber \\
&=& \negs \frac{A_{n-2}}{(n-1)} \int \limits_{0}^{\frac{N}{2}} d\eta \, \, \biggl[ \biggl( \sqrt{t_0}+
\frac{\eta}{2} \biggr)^{n} + \biggl( \sqrt{t_0}-\frac{\eta}{2} \biggr)^{n} \biggr] \biggl(
\frac{N}{2} -\eta \biggr)^{n-1}. \nonumber \\ 
&=&\negs  \frac{2 A_{n-2}}{(n-1)}  \sum\limits_{l=0}^{n-1} \sum\limits_{k=0}^{[n/2]} (-1)^l\binom{n}{2k}
\binom{n-1}{l} t_0^{\frac{n-k}{2}} \biggl(\frac{N}{2}\biggr)^{n-1-l}
\biggl(\frac{1}{2}\biggr)^{2k} \int\limits_{0}^{\frac{N}{2}} d\eta \,\,\eta^{l+2k}  \nonumber \\
\end{eqnarray} 
Using the relation Eq (\ref{binomreln}) and expanding in orders of  $N$,  this simplifies to 
\begin{eqnarray}
\volalex& =&  \frac{2 A_{n-2}}{(n-1)} \biggl( \frac{N\sqrt{t_0}}{2} \biggr)^n \biggl( \frac{1}{n} +
\frac{N^2(n-1)}{16 t_0 (n+1)(n+2)} + O(N^4) \biggr)  \nonumber \\ 
&=&\volflat  \biggl( 1+ T^2 \frac{n(n-1)}{16 t_0^2 (n+1)(n+2)} + O(N^4) \biggr)
\end{eqnarray}

Again, we see that the $O(T^3)$ terms are zero and we conclude that not only $\chi(n)=0$ but
also $\kappa(n)=0$. This suggests that although the term $\partial_0 R_{00}$ appears non-trivially
in the tangent space Jacobian $\JJ$, there is a non-trivial effect from the acceleration of the
null-geodesics which must serve to cancel this contribution\footnote{Note that the $O(T^2)$ terms in both cases  are  compatible with our expression for
the volume Eq (\ref{volgenn}) to that order.}.

\section{Conclusions} 

In this note we have shown explicitly how to calculate the volume up to $O(T^2)$ for a small  causal
diamond $\Alex[p,q]$  in
arbitrary dimensions using Myrheim's approach. Our main result is that  in any  dimension the boundary
term of Myrheim  \cite{myr}  takes the form 
\begin{equation} 
I_{\Delta\Alex[p,q]}=\volflat \, \biggl( \frac{T^2}{24} \, R_{00}(0) \biggr). 
\end{equation} 
The dimension independence of the factor multiplying $\volflat$ is intriguing. At its most mundane,
this stems from the fact that to this order it is only the {\it sum} of the eigenvalues of first
order perturbation theory that contribute. This sum is itself dimension independent since it
involves the trace $\sum_{i=1}^{n-1} R_{0i0i}(0)=\sum_{i=1}^{n} R_{0i0i}(0)=R_{00}(0)$.  Whether
there is a more profound reason for this universality is however unclear.  

Importantly, our analysis allows us to generalise this result to other integrals over the region
$\Alex[p,q]$, so that  for any integrable function $\Phi(x^\mu)$   in $\Alex[p,q]$
\begin{equation} 
\int\limits_{\Alex[p,q]} d^nx \,\, \Phi(x^\mu) =   \biggl(1+  \frac{T^2}{24} R_{00}(0) \biggr)
\int\limits_{\Alexflat} d^nz \,\,\,\widetilde \Phi(z^\mu) + O(T^{3+m})
\end{equation} 
where $m$ is the order of the flat spacetime integral.  This has useful implications for
calculations in causal set theory. Recently it has been used to find an expression for the discrete 
Ricci scalar in a causal set in terms of the abundance of ``k-chains'', a construction which differs substantially
from that previously obtained by Benincasa and Dowker \cite{bd}.

We have also extended our analysis to examine the order $O(T^3)$ contribution. Using a combination
of our analysis and the approach of Gibbons and Solodukhin \cite{gs} we find from calculations of
the spacetime volume in two types of FRW spacetimes that the $O(T^3)$ contribution to the volume in
fact vanishes. Since the spacetime volume plays such a key role in the continuum approximation of
causal set theory, we conjecture that the first non-trivial higher derivative correction to the
Einstein-Hilbert action from causal set theory must contain at least second derivatives of the
curvature (modulo boundary terms), thus distinguishing the causal set effective action from other
higher derivative theories. Our conjecture has obvious limitations since it rests solely on the
assumption that these higher order corrections obtain only from contributions to the volume. We have
neglected entirely the effect of other geometric and topological contributions that might arise in a
more subtle fashion in causal set theory, although these are at present hard to construe. To
substantiate the conjecture requires a far better understanding than we have at present of the
discrete-continuum correspondence, and in particular the precise manner in which locality is
recovered from causal set quantum gravity.

\end{document}